\documentclass[aps,showkeys,showpacs,nosuperscriptaddress,onecolumn]{revtex4}

\usepackage{psfrag}
\usepackage{graphicx}
\usepackage{amsmath}
\usepackage{hyperref}
\usepackage{color}
 
\begin{document}

\title{Evacuation in the Social Force Model is not stationary}

\author{P.~Gawro\'nski}
\email{gawron@newton.ftj.agh.edu.pl}
\affiliation{
Faculty of Physics and Applied Computer Science, AGH University of Science and Technology, al. Mickiewicza 30, PL-30059 Krak\'ow, Poland
}
\author{M.~K\"ampf}
\email{mirko.kaempf@physik.uni-halle.de}
\affiliation{
Institut f\"ur Physik, Martin-Luther-Universit\"at Halle-Wittenberg, Halle/Saale, Germany 
}
\author{J.~W.~Kantelhardt}
\email{jan.kantelhardt@physik.uni-halle.de}
\affiliation{
Institut f\"ur Physik, Martin-Luther-Universit\"at Halle-Wittenberg, Halle/Saale, Germany 
}
\author{K.~Ku{\l}akowski}
\email{kulakowski@novell.ftj.agh.edu.pl}
\affiliation{
Faculty of Physics and Applied Computer Science, AGH University of Science and Technology, al. Mickiewicza 30, PL-30059 Krak\'ow, Poland
}
 
\date{\today}
 
\begin{abstract}
An evacuation process is simulated within the Social Force Model. Thousand pedestrians are leaving a room by one exit. We investigate the stationarity of the distribution of time lags between instants when two successive pedestrians cross the exit. The exponential tail of the distribution is shown to gradually vanish. Taking fluctuations apart, the time lags decrease in time till there are only about 50 pedestrians in the room, then they start to increase. This suggests that at the last stage the flow is laminar. In the first stage, clogging events slow the evacuation down. As they are more likely for larger crowds, the flow is not stationary. The data are investigated with detrended fluctuation analysis and return interval statistics, and no phase transition is found between the stages of the process.

 \end{abstract}
 
 

\maketitle
  
%% ###########################################################################
\section{\label{sec-intro} Introduction}
%% ###########################################################################

A human crowd is a specific system, which is of interest for various specialists for different reasons. A physicist is willing to treat a crowd as a gas or a fluid of interacting particles \cite{dh0}, a psychologist can concentrate on the process of self-categorization in crowds \cite{rei}, and a sociologist 
asks for emergence of norms in a crowd \cite{tuk}. In an interdisciplinary approach, these perspectives overlap. In the Social Force Model (SFM), designed by Dirk Helbing and coworkers \cite{dh1,dh2} in the 90's, physical interactions are combined with action of the social norm of keeping distance to unknown persons \cite{hall}. Although the differential equations used there can be considered as computationally complex, the SFM seems to describe properly
the collective effects in crowd, which appear in particular in emergency situations, as an evacuation. In simpler techniques, such as cellular automata or lattice gas models, a prescribed area is reserved for each pedestrian, and the influence of other pedestrians is reduced to short range interactions. Even if some specific effects such as clogging and arching are reproduced (as, for example, in \cite{sc,ex}), dynamics of a pedestrian in these techniques is fully determined by her/his local environment. On the contrary, in most of crowd disasters the crowd size was essential \cite{r1}. For reviews on the SFM and other techniques and a list of literature we refer to \cite{r1,r2,r3,schad,xi}.

Here we are interested in forces exerted by masses of pedestrians, when physical interactions accumulate and the crowd size does matter. Namely, we intend to investigate how the number of pedestrians in a room influences the flow through an exit. Therefore we designed a numerical experiment as follows. A number of pedestrians is waiting, crowded, at the exit. At $t=0$, the exit is opened and the crowd is pushing towards it. Let us denote the number of pedestrians who remain in the room at time $t$ as $n(t)$. In this setup, an experiment with $N$ pedestrians provides data on experiments 
for all $n<N$, because there is no stage when people gather at the exit. Provided that the flow at the exit depends on the number of pedestrians in the room, we should observe this dependence by just measuring the curve $n(t)$. Alternatively, we can measure the time gaps between successive crossings of exit. One can write
\begin{equation}
n(t)=N-\sum_{i=1}^N\Theta(t-t_i)
\end{equation}
where $t_i$ is the time instant when the number of pedestrians in the room changes from $i+1$ to $i$. We are going to concentrate on the series of the time lags $\Delta_i=t_{i+1}-t_{i}$. Is it stationary? Are there long-term correlations and/or regimes with characteristically different behavior?

Up to our knowledge, this question was not posed directly in the literature, but the shape of the function $n(t)$ was obtained several times by different authors. In the next section we gather the results obtained by other authors which are directly close to our specific interest. For reasons explained above, we do not refer to simulations done with the cellular automata and lattice gas model. In the third section the SFM is briefly described. In Section IV we describe the technique of the data analysis. Finally, we show our numerical results (Section V) and discuss them (section VI).

%% ###########################################################################
\section{\label{sec-others} What is known}
%% ###########################################################################

In \cite{dh1}, the model equation of motions of pedestrians were formulated. Among other results, an effect of pressure of crowd was demonstrated; out of two groups attempting to cross the door in opposite directions, the larger group was prevailing until the larger group became smaller. In this paper, a noise term is included to the equation of motion, hence the term 'Langevin equations'. In \cite{dh2}, the same SFM equations were used without the noise term. There, the desired velocity was associated with the level of panic. Also, the evacuation time dependence on the desired velocity was found to display a minimum. Also, when the desired velocity was increased, a change of the process from a laminar to a clogging mode of the crowd behavior was observed. As the authors remarked, the effect was less pronounced for wider exits. Note that, as noted in \cite{muir}, the experimental data collected in planes do not show abrupt changes of the effectivity of evacuation when the exit width is changed from 0.6 m to 1.8 m. As shown in \cite{schad} with more experimental data, the relation between the bottleneck width and the flow of pedestrians does not show any threshold.

In \cite{pd1}, three curves are shown, obtained by simulations with using the SFM, on the number of pedestrian who left the room against time. The curves were obtained for 200 pedestrians and three values of the desired velocity: 0.8, 2.0 and 6.0 m/s. First curve (0.8) shows that at the last stage of evacuation the flow decreases. This effect exists also, but is weaker, for the second curve (2.0), but not for the third one (6.0). Instead, the latter curve was found to be particularly noisy. In Fig. 3 of \cite{pd1}, the distribution of clogging delays is shown for up to 160 people and the three above given values of the desired velocity. Each curve shows a clear maximum between 0.2 and 0.4 s. In this and subsequent paper \cite{pd2}, the cluster size distribution is also investigated, where a cluster means a group of people in physical contact between them. For the laminar and the turbulent flow, this distribution is found to be qualitatively different.

In \cite{imps}, the formalism of optimization, developed by authors for other purposes, has been applied to the evacuation problem. Both clogging and arching have been observed in the simulation. The evacuation time dependence on the number of people was found to decrease in a non-linear way, but no minimum of this curve was found. In \cite{rk}, the influence is investigated of the desired velocity on the evacuation time, the latter being a measure of panic. The mode of motion when the evacuation time increases with the desired velocity has been classified as turbulent. The effect of wider exit was investigated directly: the desired velocity where the evacuation time displays a minimum was shifted towards larger value with the exit width. In \cite{sey} and references cited therein, an experiment performed in a wardroom with a group of 70 soldiers is described. It was found among other results that the clogging is more likely if the number of persons is larger than 45.

For completeness we remark also our two recent papers \cite{my1,my2}, where the SFM was used to investigate the chances that persons in the crowd can decide about themselves. The stationarity of the process of evacuation was not investigated there.

%% ###########################################################################
\section{\label{sec-model} The model and simulation}
%% ###########################################################################

The simulation is based on the model of crowd dynamics, described by Helbing et al. \cite{dh2}. In this model, the equation of motion 
of a person $i$ of mass $m$ is written as
\begin{equation}
m\frac{d\mathbf{v}_i}{dt}=m\frac{\mathbf{v}(\mathbf{r}_i)-\mathbf{v}_i}{\tau}+\sum_{j(\ne i)}\mathbf{f}_{ij}+\sum_W\mathbf{f}_{iW}
\end{equation}
where the first term on the right hand side is the tentative acceleration of a person $i$ who intends to have the velocity $\mathbf{v}(\mathbf{r}_i)$,
(its length commonly termed as the desired velocity) dependent on the coordinates $\mathbf{r}_i$. As a rule, the vector $\mathbf{v}$ points to the exit center (large distance from the person to the exit) 
or to the closest point of the exit (small distance). Further, $\tau$ is the characteristic time of this acceleration, $\mathbf{v}_i$ is the actual velocity of $i$-th person, $\mathbf{f}_{ij}$ is the 
force exerted on $i$-th person by $j$-th person, and $\mathbf{f}_{iW}$ is the force exerted on $i$-th person by a wall $W$. The force
$\mathbf{f}_{ij}$ contains three components; 'social' interaction which describes the tendency of $i$ and $j$ to keep distance between 
each other, and two physical interactions between their bodies: radial force and slide friction. The social part of interaction is also adapted from \cite{dh2}. It is given by
\begin{equation}
f_{ij}^{psych}=A_i\exp((2R-\big\|\mathbf{r}_i-\mathbf{r}_j\big\|)/B) 
\end{equation}
where $A_i$ and $B$ are constants, $R$ is the mean 'radius' of the vertical projection of the human body, and $\mathbf{r}_i$ is the position of $i$-th agent. The parameters of the simulation are adapted from \cite{dh2}: the amplitude of the social force $A_i=2000 N$, the constant $B$ which is responsible for the spatial dependence of the social force is $0.08 m$, the radii of agents $R=0.3 m$, their masses $m=75 kg$, the characteristic time of acceleration is $\tau=0.5 s$ and the absolute value of the desired velocity is $\vert \mathbf{v} \vert = 3 m/s$. As remarked in \cite{dh2}, these values allow to reproduce experimental interpersonal distances and flows through bottlenecks. Also, the same values are assumed for all persons, to minimize the number of parameters. The instant values of the velocities $\mathbf{v}_i$ allow to update the positions $\mathbf{r}_i$ as well. The equations of motion are solved with the Runge-Kutta method of 4-th order.

The simulation was performed as follows. $N=10^3$ pedestrians were gathered at a the closed exit of width of 1 m, which was opened at $t=0$. The time lags $\Delta_i=t_{i+1}-t_{i}$ were measured between crossing of the exit by subsequent pedestrians. The simulation was repeated 100 times. 

%% ###########################################################################
\section{\label{sec-ana} Data analysis}
%% ###########################################################################

In our analysis procedure, we split the data of each run into ten non-overlapping parts corresponding to 100 persons leaving the room.  Each part is analyzed independently, but averages over all 100 simulation runs are calculated to improve statistics.  

Quantitatively, correlations between time lags $\Delta_i$ separated by $s$ people are defined by the (auto-) correlation function, 
\begin{equation}
C(s) \equiv {\frac{1}{L-s}}\sum_{i=0}^{L-s-1} (\Delta_i - \bar \Delta) (\Delta_{i+s} - \bar \Delta),
\label{autocor} \end{equation}
where $L$ is the length of the considered data part and $\bar \Delta$ is the average time lag in this part. If the time lags are uncorrelated, $C(s)$ is zero for $s$ positive. If correlations exist up to a certain number of people $s_\times$, the correlation function will be positive up to $s_\times$ and vanish above $s_\times$. For the relevant case of long-range correlations, $C(s)$ decays as a power law, 
\begin{equation}
C(s) \sim s^{-\gamma}, \quad 0<\gamma<1.  \label{gamma}
\end{equation}
A direct calculation of $C(s)$ is hindered by the non-stationarities and trends in the data, since $\bar \Delta$ is not constant.  We thus apply return interval statistics and detrended fluctuation analysis to study short-term and long-term correlations in the data, respectively.

Usually, return interval statistics (RIS) study the time intervals between 'extreme events' that exceed a given threshold \cite{Bunde.2003,Bunde.2005,Altmann.2005,Eichner.2007,Santhanam.2008,Moloney.2009}.  In a sequence of uncorrelated values ('white noise'), these return intervals are also uncorrelated and distributed according to a Poisson distribution, 
\begin{equation}
P_{q}(r)=(1/R_{q})\exp (-r/R_{q}),  \label{simple}
\end{equation}
where $R_{q}$ is the mean return interval $\langle r \rangle$ for the given threshold $q$. For long-correlated data, on the other hand, a stretched exponential distribution 
\begin{equation}
P_{q}(r)={\frac{a_{\gamma }}{R_{q}}}\exp [-b_{\gamma }(r/R_{q})^{\gamma }] \label{stretched}
\end{equation}
has been observed \cite{Bunde.2003,Bunde.2005,Altmann.2005,Eichner.2007}, where the exponent $\gamma$ is the correlation exponent from Eq.~(\ref{gamma}), and the parameters $a_{\gamma }$ and $b_{\gamma }$ are independent of $q$ \cite{Altmann.2005,Eichner.2007}. If, on the other hand, the data is nearly deterministic (and not random), all return intervals will fluctuate weakly around the typical value $R_q$, giving rise to, e.~g., a Gaussian distribution,
\begin{equation}
P_{q}(r)=\frac{1}{\sigma \sqrt{2 \pi}} \exp[- (r - R_{q})^2 /(2 \sigma)],  \label{gauss}
\end{equation}
for $r>0$ with the small standard deviation $\sigma \ll R_q$.

Here we consider each event of a person leaving the room as an extreme event, so that the return intervals $r$ are identical with the time lags $\Delta_i$, and $R_q$ is identical with $\bar \Delta$.  There is thus no threshold $q$ for extreme events, but we get much more statistics.  Our RIS focus on short-term correlations (between successive persons).   

Detrended fluctuation analysis (DFA) \cite{Peng.1994} has become a widely-used technique for the detection of long-range correlations in noisy, nonstationary time series \cite{Kantelhardt.2001,Hu.2001,Chen.2002}. The DFA procedure consists of four steps. First we determine the 'profiles' $Y(j)\equiv \sum_{i=0}^{j} (\Delta_i- \bar \Delta)$, $j=1,\ldots ,L$. Secondly, we divide $Y(j)$ into $L_{s}\equiv \mathrm{int}(L/s)$ non-overlapping segments of equal length $s$. Thirdly, we calculate the local trend for each segment by a least-square fit of the data. Linear, quadratic, cubic, or higher order polynomials can be used in the fitting procedure (conventionally called DFA1, DFA2, DFA3, $\ldots $) \cite{Bunde.2000}. Then we determine the variance $F_{s}^{2}(\nu)$ of the differences between profile and fit in each segment $\nu$.  Fourthly, we average $F_{s}^{2}(\nu)$ over all segments and take the square root to obtain the fluctuation function $F(s)$. Since we are interested in how $F(s)$ depends on the time scale $s$, we have to repeat steps 2 to 4 for several $s$. Apparently, $F(s)$ increases with increasing $s$. If data $\Delta_i$ are long-range power-law correlated according to Eq.~(\ref{gamma}), $F(s)$ increases, for large values of $s$, as a power-law,
\begin{equation}
F(s)\sim s^{\alpha },\quad \alpha =1-\gamma /2.  \label{alpha}
\end{equation}
To determine the asymptotic scaling behavior of this fluctuation function we plot $F(s)$ as a function of $s$ on double logarithmic scales and calculate the slope $\alpha$ by a linear fit in the regime $10< s <100$. This way, short-term correlations affecting less than 10 persons subsequently exiting the room are ignored in the analysis.

%% ###########################################################################
\section{\label{sec-res} Results}
%% ###########################################################################

\begin{figure}[ht]
{\centering \includegraphics[width=12cm]{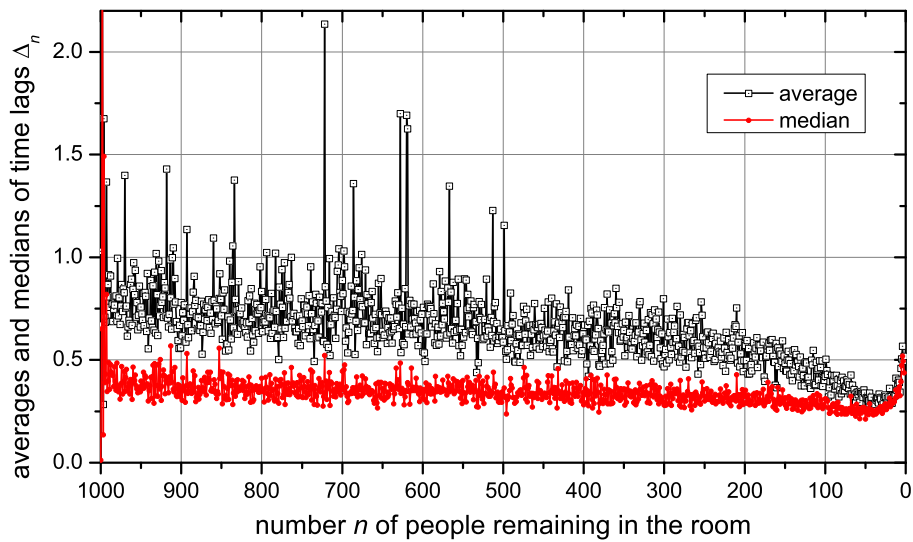}}
\caption{Mean values and medians of the time lags $\Delta_n$ against the number $n$ of pedestrians remaining in the room. Data from 100 simulations with $N=1000$ people are included. The strong fluctuations at the beginning of the simulations ($n$ close to 1000) are transient effects due to the opening of the exit. One can distinguish two regimes (approximately for $n>50$ and $n<50$) when comparing averages and medians.} \label{fig-1}
\end{figure}

In Fig.~1, results are shown for the average values and medians of $\Delta_n$ against the number $n$ of pedestrians remaining in the room. As we see, two parts of the process can be distinguished. In the first regime from $n = 990$ to $\approx 50$, averages and medians differ significantly and the fluctuations of $\Delta_n$ are rather strong. However, the fluctuations seem to  decrease gradually, and so do the measured values. In the second regime from $n \approx 50$ to 0, where fluctuations are small, averages and medians are nearly identical, and the measured values increase slightly.  This change of behavior can be interpreted as a cross-over from a stage with temporal cloggings to a laminar stage. In the latter case, the crowd behind pedestrians at the exit is large enough to push them out but not large enough to cause clogging. The appearance of the second stage of evacuation agrees with the character of two out of three discharge curves, presented in Fig.~2 of \cite{pd1}, for small and moderate desired velocity.

\begin{figure}[ht]
{\centering \includegraphics[width=15cm]{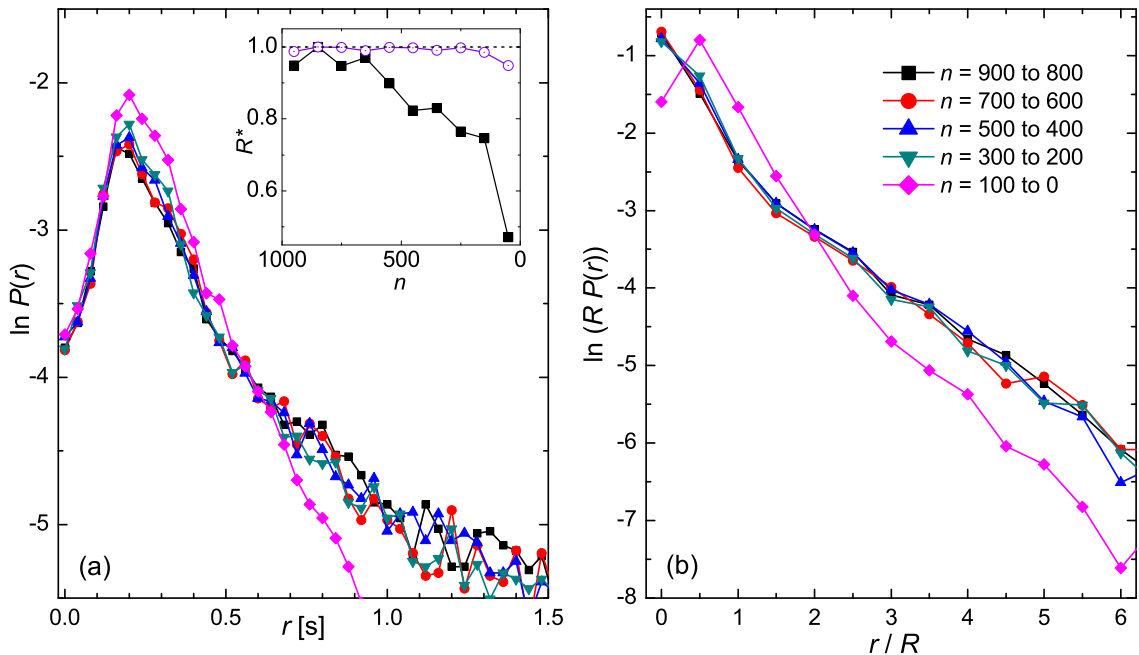}}
\caption{(a) Distributions $P(r)$ and (b) scaled distributions $R \cdot P(r/R)$ of the time lags $r=\Delta_i$ with mean $R=\bar \Delta$ for $n=900$ to 800 (black squares), $n=700$ to 600 (red circles), $n=500$ to 400 (blue triangles up), $n=300$ to 200 (green triangles down), and $n=100$ to 0 (violet diamonds) persons remaining in the evacuated room. The unscaled distributions in (a) show that the nearly Gaussian peak for short time lags is hardly changing, while the exponential tail is decaying with decreasing $n$.  The inset in (a) shows the fitted slopes $R^*$ of the exponentials decays (black squares together with the Pearson correlation coefficients of the fits (blue open circles).  The scaled distributions in (b) show that $R=\bar \Delta$ characterizes the exponential peak fairly well.} \label{fig-2}
\end{figure}

In Fig. 2(a), we show the distributions of return intervals $r=\Delta_n$, i.~e. $P(r)$ gathered in shorter parts of the data, where the departure from stationary flow can be approximately neglected.  However, comparing the statistics, we see the differences. Initially, for large numbers $n$ of people in the room, there is a large exponential tail of the distribution, formed by the clogging events and corresponding to the Poisson distribution Eq.~(\ref{simple}) with a modified prefactor.  In addition, there is a distinguished maximum near the time lag $r = \Delta_n \approx 0.2$ s, which is approximately described by a Gaussian distribution Eq.~(\ref{gauss}), also with a modified prefactor.  The center of this peak agrees approximately with the results shown in Fig.~3 of \cite{pd1}.  We see that there are apparently two distinct components in the time lag distributions: typical short time lags around 0.2 s (probably due to persons successively exiting without delays) and exponentially distributed longer time lags (probably due to interruptions in the flow of exiting people because of clogging or arching effects).  

For smaller numbers $n$ of people in the room the exponential tail decreases, to nearly vanish during the last stage, i.~e. for $n=50$ to 0.  Simultaneously, the nearly Gaussian peak for short time lags is hardly changing.  While the behavior of the short time lags is well characterized by the nearly constant median of $\Delta_n$ (see Fig.~1), the changing averages of $\Delta_n$, i.~e. $\bar \Delta$, characterize the exponential behavior for long time lags.  This is confirmed by the plot of scaled distributions shown in Fig.~2(b).

The application of DFA to the data from each of the ten parts yields fluctuation exponents $\alpha$ very close to 0.5, which proves the absence of relevant long-term correlations during all stages of the evacuation procedure.  Specifically, we obtain $\alpha = 0.55$ and 0.54 for the first two parts (between 1000 and 800 people in the room) and values between 0.50 and 0.53 for all other parts.  Since the systematic error of such fluctuation exponents is around 0.05 for time series of just 100 values \cite{Kantelhardt.2001}, all of these numerical results are fully consistent with the null hypothesis of only short-term correlations in the data.

\begin{figure}[ht]
 {\centering \includegraphics[width=12cm]{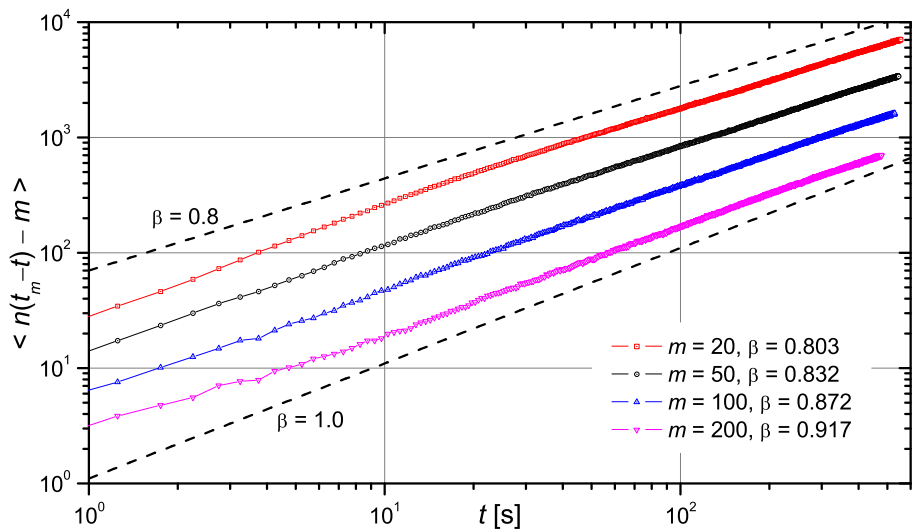}}
\caption{The average discharge curve $\langle n(t_m - t) \rangle - m$ is shown versus the time $t$ that measures the time interval till only $m$ persons remain in the room.  Curves are shown for $m=20$ (red), 50 (black), 100 (blue), and 200 (violet).
Each curve is fitted with $t^{\beta}$ in the scaling regime $50<t/s<500$.} \label{fig-3}
\end{figure}

Finally, we are interested in the scaling behavior of the so-called discharge curve, i.~e.~the number of people remaining in the room against time.  However, as we observe in Fig.~1, for the last approximately 50 persons the character of the curve changes. Then, for each numerical experiment $j=1,...,100$ we determine the time $t^j_{50}$ when $m=50$ persons are left in the room in $j-th$ experiment, and we investigate the dependence of $n(t^j_{50}-t) - 50$ on $t$ for $t<t^j_{50}$.  Again, $n(t)$ is the number of pedestrians in the room.  This dependence is averaged over all 100 experiments.  The result is shown in Fig.~3 in the log-log scale.  Fitting the result to a straight line we get an exponent $\beta$, which indicates if and how the evacuation speed depends on the crowd size in the first stage of evacuation.  A result $\beta=1$ means lack of this dependence; if $\beta=1$, the evacuation is stationary at least in its first stage.  However, our result is that clogging events make the evacuation slower, and these events are more likely if the number of pedestrians in the room is larger.  The latter effect agrees with the experimental result in \cite{sey}.  

Effectively, our exponent $\beta$ is smaller than one; we get $\beta=0.832 \pm 0.001$ for the fitting range $50$s$ < t < 500$s and parameter $m=50$ as cutoff point of minimal number of persons in the room.  Fig.~3 also shows that the numerical value of $\beta$ is depending on $m$, reaching a bit smaller values for smaller $m$ and larger values for larger $m$.  We admit that perhaps the scaling regime is not fully reached yet because the crowd may be too small.  Still, the conclusion that $\beta$ is close to 0.85 seems to be well grounded.  This form of the scaling relation allows to extrapolate the results for larger crowds.

%% ###########################################################################
\section{\label{sec-dis} Discussion}
%% ###########################################################################

The results indicate that the probability distribution of the time lags $\Delta_i$ changes in time. In other words, the evacuation process simulated here is not stationary. One of the consequences is that the total evacuation time depends on the number $n$ of pedestrians in the room, and therefore it is not a good measure of the efficiency of the process. If the size of the crowd is large, effects of clogging appear which are absent for small numbers of pedestrians. Although the case of larger desired velocity is not investigated here, we can reasonably expect that our findings will be particularly important if large desired velocity happens to be combined with large crowd. Effects of victims, who become obstacles, can only enhance the nonstationary character of the process. 

A straightforward interpretation of our result is that the SFM successfully describes the effect of cumulation of the physical forces between agents at the exit. Keeping the hydrodynamic analogy, the pressure at the exit increases with the crowd size. If this pressure exceeds some critical value, pedestrians at the exit are not able to move, even if they are close to the exit. This is the origin of the observed large values of the time lags $\Delta_i$, and the dependence of the size of the exponential tail of the lag distribution on the size of the crowd.

On the other hand, the data analysis with the return interval statistics and detrended fluctuation analysis shows that there is no phase transition between the first stage, when the mean time lag $\bar \Delta$ decreases, and the second stage, when the mean time lag increases.  Also, correlations between pedestrians leaving the room in subsequent times are broken by the clogging events.  As a result, the observed tail of the probability distribution of $\Delta$ is Poissonian.

\begin{acknowledgments}
The research is partially supported by the European Union within the FP7 project SOCIONICAL, No. 231288.
\end{acknowledgments}

\end{document}